\documentclass[twocolumns]{aa}
\usepackage{graphicx}
\usepackage{txfonts}
\usepackage{psfig}
\usepackage{epsfig}

\def \deg{^\circ}

\def \hcm {\hbox {\ifmmode $ cm$^{-2}\else cm$^{-2}$\fi}}

\def\approxgt{\mathrel{\hbox{\rlap{\lower.55ex \hbox {$\sim$}}
        \kern-.3em \raise.4ex \hbox{$>$}}}}
\def\approxlt{\mathrel{\hbox{\rlap{\lower.55ex \hbox {$\sim$}}
        \kern-.3em \raise.4ex \hbox{$<$}}}}

\begin{document}

\title{The luminous host galaxies of high redshift BL Lac objects.}

\author{J.K. Kotilainen\inst{1}, T. Hyv\"onen\inst{1} \and R. Falomo\inst{2}}

\offprints{J.K. Kotilainen}

\institute{Tuorla Observatory, University of Turku, V\"ais\"al\"antie 20, 
FIN--21500 Piikki\"o, Finland\\
\email{jarkot@utu.fi; totahy@utu.fi}
\and
INAF -- Osservatorio Astronomico di Padova, Vicolo dell'Osservatorio 5, 
35122 Padova, Italy\\
\email{falomo@pd.astro.it}
}

\date{Received; accepted }

\abstract{
We present the first near-infrared $Ks$-band (2.1 $\mu$m) imaging study of 
a sizeable sample of 13 high redshift 
(0.6 $<$ z $<$ 1.3) BL Lac objects in order to characterize the properties of 
their host galaxies. We are able to clearly detect the surrounding nebulosity 
in eight objects, and marginally in three others. In all the well resolved 
objects, we find that the host galaxy is well represented by 
a de Vaucouleurs $r^{1/4}$ 
surface brightness law. In only two cases the object remains unresolved.  
 These new observations  
represent in most cases the first  detection of the host galaxy and 
taken together with previous optical  studies of  z $>$ 0.5 BL Lacs 
substantially increase the number of detected  hosts (from $\sim$20 
to $\sim$30). This dataset allows us to explore the evolution of BL Lac hosts 
from z $\sim$1 to the present epoch.

We find that the host galaxies of high redshift BL Lacs are large 
(average bulge scale length $<$R(e)$>$ $\sim$ 7 kpc) and similar to 
those hosting low redshift BL Lacs, indicating that there is no evolution 
in the host galaxy size. On the other hand, these host galaxies are 
very luminous (average $<$M(K)$>$ = -27.9$\pm$0.7). They are $\sim$3 mag 
brighter than the typical galaxy luminosity L*, and $\sim$1-1.5 mag 
more luminous than brightest cluster galaxies at low redshift. They are also 
$\sim$1 magnitude brighter than radio galaxies at low redshift and they appear 
to deviate from the $K$--z relationship of radio galaxies. On the other hand, 
these high luminosities agree with the few optical studies of high redshift 
BL Lacs and are similar to those of flat spectrum radio quasars studied by us 
in the near-infrared. 

The nuclear luminosity and the nucleus--galaxy luminosity ratio of 
the high redshift BL Lacs are much larger than those found for  low redshift 
BL Lacs and  similar to those observed in  flat spectrum radio quasars at 
similar redshift. 
This mainly reflects the selection effects in the surveys and may be 
due to either an higher intrinsic nuclear luminosity, or due to 
enanched luminosity because of strong beaming. 
Contrary to what is observed in low redshift BL Lacs, the luminosities of 
the host galaxy and of the nucleus appear fairly well correlated, 
as expected from the black hole mass - bulge luminosity relationship 
found in nearby spheroids, if the nuclear emission works at the same regime.  
Our observations indicate that high redshift BL Lacs radiate with a wide range 
of power with respect to their Eddington luminosity, and this power is 
intermediate between the low level observed in nearby BL Lacs and 
the higher level occurring in luminous radio-loud quasars.

The comparison with BL Lac host galaxies at lower redshift suggests that 
there is a $\sim$2 mag brightening of the hosts. 
We argue that the large luminosity of the hosts is due to 
a strong selection effect in the surveys of BL Lacs that makes observable only 
the most luminous sources at z $>$ 0.5 and produces a correlation between 
the nuclear and the host luminosity that emerges at high redshift.
However, this may also suggest a strong luminosity evolution which is 
inconsistent with a simple passive evolution of the stars in 
the host galaxies, and requires a contribution from recent star formation 
episodes that takes place at z $>$ 0.5. 

\keywords{BL Lacertae objects:general -- Galaxies:active -- 
Galaxies: elliptical and lenticular. cD -- Galaxies:nuclei -- 
Galaxies:photometry -- Infrared:galaxies}
}

\titlerunning{Evolution of BL Lac hosts}
\authorrunning{Kotilainen et al.}

\maketitle

\section{Introduction}

Models for the formation of galaxies based on hierarchical merging 
of galaxies  (e.g. Franceschini et al. 1999; 
Kauffmann \& H\"ahnelt 2000; Di Matteo et al. 2003) predict 
a close link between the cosmological formation and evolution of 
massive spheroids and the processes that fuel their central 
black holes (BH). This connection is strongly supported by 
the virtually ubiquitous detection of supermassive BHs in the centers 
of nearby inactive elliptical and bulge-dominated spiral galaxies 
(e.g. Magorrian et al. 1998; van der Marel 1999), 
and the strong correlation between the BH mass and the luminosity (mass) 
and kinematics (velocity dispersion) of the host galaxy 
(e.g. Magorrian et al. 1998; Gebhardt et al. 2000; 
Ferrarese \& Merritt 2000; McLure \& Dunlop 2002; Marconi \& Hunt 2003). 
More circumstantial evidence is provided by the census of local 
BH masses being comparable to the BH mass integrated over 
the Hubble time using the quasar luminosity function 
(Yu \& Tremaine 2002), and by the similarity of the strong 
cosmological evolution of the AGN population 
(e.g. Dunlop \& Peacock 1990; Warren, Hewett \& Osmer 1994) 
to the evolution of the star formation history of the universe 
(e.g. Madau, Pozzetti \& Dickinson 1998; Franceschini et al. 1999). 
It is thus plausible that galaxy bulge formation simultaneously 
triggers star formation and accretion onto the BH. 
If the nuclear activity is a common transient phenomenon during 
the lifetime of all giant galaxies (Cavaliere \& Padovani 1989), with one 
or several phases of nuclear activity with recurrent 
accretion episodes, understanding the luminosity evolution of AGN 
host galaxies is fundamental to study the link between AGN activity and 
galaxy formation and evolution (e.g. Franceschini et al. 1999).

Detailed imaging studies of nearby AGN have shown that the host galaxies 
of low redshift (z $<$ 0.5) powerful radio-loud AGN 
(radio-loud quasars [RLQ] and FR II radio galaxies [RG]) are luminous 
large elliptical galaxies (e.g. Dunlop et al. 2003; 
Falomo et al. 2003; Pagani, Falomo \& Treves 2003) that have 
the same structural and photometric properties (e.g. the same 
Kormendy relation) and appear to follow the same BH mass - host 
luminosity relation as that found for inactive nearby galaxies 
(McLure \& Dunlop 2002). At higher redshift (z $>$1), the host galaxies 
are more luminous but still follow a similar Kormendy relation and 
are consistent with having formed at very high redshift (z $\geq$ 2) 
followed by passive stellar evolution of their host galaxies 
(e.g. McLure \& Dunlop 2000; Kukula et al. 2001; Zirm, Dickinson \& Dey 2003; 
Falomo et al. 2004). Passive evolution is also supported by the few 
available spectroscopic studies of low redshift quasar hosts and RGs 
(e.g. Nolan et al. 2001), indicating that they are dominated by an old 
evolved stellar population. This behaviour is different from 
brightest cluster galaxies (BCG; Aragon-Salamanca, Baugh \& Kauffmann 1998) 
and normal inactive elliptical galaxies 
(e.g. Stanford, Eisenhardt \& Dickinson 1998; 
Bell et al. 2004), which exhibit flat luminosity evolution from z $\sim$1. 
However, it has for long been suggested that the maximum AGN power 
increases with the host mass (e.g. Smith et al. 1986; 
O'Dowd, Urry \& Scarpa 2002; Dunlop et al. 2003), resulting in 
apparent evolution in flux-limited samples. Furthermore, the amount 
of scattered nuclear light and the level of extended line emission 
from ionized gas around the nucleus increases with the AGN power. 
This may contaminate the measurement of the host galaxy emission and, 
if not taken into account, may lead to an erroneous evaluation of 
the host luminosity and its trend with redshift.

Less is known about the evolution of the host galaxies of 
low luminosity radio-loud AGN (FR I RGs and BL Lac objects). In agreement 
with luminous radio-loud AGN, low redshift FR I RGs are large 
luminous ellipticals which follow the same Fundamental Plane defined by 
the properties of inactive ellipticals (Govoni et al. 2000; 
Bettoni et al. 2003). FR I RGs are, however, difficult to detect at 
z $>$ 0.2 (Magliocchetti, Celotti \& Danese 2002) 
where the evolutionary effects start to become important. 
The only reasonable possibility to study the host galaxies of 
low luminosity radio-loud AGN over a wide redshift range is therefore 
afforded by BL Lac objects, 
which together with flat spectrum radio quasars (FSRQ) form the AGN 
subclass of blazars. However, unlike BL Lacs which have weak or absent 
optical line emission (e.g. Scarpa \& Falomo 1997), FSRQs have strong 
emission lines of similar intensity to "normal" 
steep spectrum radio quasars (SSRQ).  

In the current unified models of AGN (Urry \& Padovani 1995), 
the parent population of FSRQs are the high luminosity lobe-dominated 
FR II RGs, whereas BL Lacs are low luminosity core-dominated FR I RGs whose 
apparent luminosity is dominated by a relativistically beamed synchrotron jet 
seen aligned close to our line-of-sight. BL Lacs are, therefore, 
intrinsically low power AGN, with much lower accretion rates than in RLQs 
(O'Dowd et al. 2002) which, due to the beaming, can be detected in 
significant numbers to high redshift. 
The nuclei of the BL Lacs are on average fainter than those of quasars, 
which makes the determination of their host properties much easier.

Optical imaging studies (e.g. Falomo 1996; Wurtz, Stocke \& Yee 1996; 
Falomo \& Kotilainen 1999; Scarpa et al. 2000a; Urry et al. 2000; 
Pursimo et al. 2002; Nilsson et al. 2003) and, more recently, 
near-infrared (NIR) imaging studies 
(Kotilainen, Falomo \& Scarpa 1998b [KFS98]; Scarpa et al. 2000b [S00]; 
Cheung et al. 2003 [C03]; Kotilainen \& Falomo 2004 [KF04]) have shown 
that virtually all nearby (z $<$ 0.5) BL Lacs are hosted by luminous, 
mostly unperturbed giant elliptical galaxies located in poor environments 
and following the Kormendy relation for inactive ellipticals. 
Their estimated BH masses (Barth, Ho \& Sargent 2003; 
Falomo, Kotilainen \& Treves 2002; Falomo et al. 2003; 
Woo et al. 2004) are in the range found for quasars, and their 
host galaxies follow the Fundamental Plane of normal ellipticals. 
The host galaxies and the environments of BL Lacs are therefore similar 
to those of FR I RGs. Possible contribution to the parent population 
from higher luminosity FR II RGs is, however, not excluded 
(e.g. Falomo \& Kotilainen 1999; Cassaro et al. 1999). 

At higher redshift (z $>$ 0.5), BL Lac hosts have been very 
little investigated, because of the increasing difficulty of both 
resolving them and the rapid cosmological (1+z)$^4$ dimming 
(i.e. by $\sim$3 mag from z = 0 to z = 1) of the host galaxy 
surface brightness. Until now, only $\sim$20 
high redshift (z $>$ 0.5) BL Lac host galaxies have been resolved, and all 
of them in the optical (including nine resolved BL Lacs in the $I$-band 
study of Heidt et al. 2004 [H04] and five in the $R$-band HST study of 
O'Dowd \& Urry 2004 [OU04]). This lack of available data has hampered until 
now a reliable study of the evolution of the host galaxies. 

In order to better investigate the evolution of the host galaxies of BL Lacs,
we have thus collected deep NIR $Ks$-band (2.1 $\mu$m) images of BL Lac 
hosts at high redshift. At variance with previous studies, the choice of 
a NIR wavelength allows us to explore the host galaxies in 
rest-frame wavelength 0.9 - 1.3 $\mu$m, where the host galaxies are 
dominated by the massive old stellar population, 
the brightness contrast between the nucleus and the host galaxy 
is significantly lower than that at shorter wavelengths, 
K--correction is relatively small and the contributions from nuclear 
scattered light and extended line emission are insignificant. Due to 
their extreme properties (especially the absence or weakness of 
emission lines), currently there exist relatively few (only $\sim$35) 
BL Lacs with a confirmed redshift z $>$ 0.5. The observed BL Lacs 
were selected from among these, with the only additional observability 
constraint of DEC $>$ -20$\deg$. 
We plan in the near future to extend this sample as the number of BL Lacs with 
known redshift at z $>$ 0.5 is expected to increase 
(e.g. Sbarufatti et al. 2005). 

\begin{table*}
\caption{The sample and journal of observations.$^{\mathrm{a}}$}
\label{journal}
\begin{flushleft}
\begin{tabular}{llllll}
\hline
\noalign{\smallskip}
Name & z & V & Date & T(exp) & FWHM \\
& & & & min & arcsec\\
(1) & (2) & (3) & (4) & (5) & (6) \\
\noalign{\smallskip}
\hline
\noalign{\smallskip}
PKS 0138-097   & 0.733  & 18.0 & 19+20/07/2002    & 65 & 0.65\\
PKS 0235+164   & 0.940  & 15.5 & 08/08/2003       & 60 & 0.85\\
B2 1308+326    & 0.997  & 19.0 & 18+19/07/2002    & 57 & 0.80\\
RXJ 14226+5801 & 0.638  & 19.0 & 19/07/2002       & 37 & 0.75\\
1ES 1517+656   & 0.702: & 15.5 & 19+20/07/2002    & 30 & 0.60\\
1ES 1533+535   & 0.890: & 17.6 & 20/07/2002       & 51 & 0.60\\
PKS 1538+149   & 0.605  & 17.2 & 08+09/08/2003    & 60 & 0.60\\
S4 1749+701    & 0.770  & 16.5 & 19/07/2002       & 60 & 0.70\\
S5 1803+784    & 0.684  & 16.4 & 08+09/08/2003    & 60 & 0.65\\
S4 1823+568    & 0.664  & 18.4 & 07/08/2003       & 46 & 0.80\\
PKS 2032+107   & 0.601  & 18.6 & 20/07/2002       & 48 & 0.55\\
PKS 2131-021   & 1.285  & 19.0 & 08+16/08/2003    & 78 & 1.05\\
PKS 2207+020   & 0.976  & 19.0 & 09+16/08/2003    & 45 & 0.80\\
\noalign{\smallskip}
\hline
\end{tabular}
\end{flushleft}
\begin{list}{}{}
\item[$^{\mathrm{a}}$] 
Column (1) gives the name of the object; 
(2) the redshift; (3) $V$-band apparent magnitude; 
(4) the dates of observation; (5) total exposure time; and (6) seeing FWHM.
\end{list}
\end{table*}

This work presents the first NIR study of 13 high redshift BL Lac 
host galaxies. The new NIR data are combined with previous optical studies 
of BL Lacs (H04, OU04) at z $>$ 0.5, in order to probe 
the cosmological evolution of the BL Lac host properties from z $\sim$1 to 
z = 0 from comparison with the hosts of lower redshift BL Lacs, 
previously studied in the NIR (KF04 and references therein); to compare 
the properties of the BL Lac hosts with those of other blazars (FSRQs) 
and ``normal'' SSRQs in the same redshift range 
(Kotilainen, Falomo \& Scarpa 1998a; Kotilainen \& Falomo 2000); to study 
the role of the host galaxy in different types of radio-loud AGN; 
and to investigate the relationship between nuclear and host luminosity in 
BL Lacs. The properties of the observed objects are given in 
Table ~\ref{journal}. In section 2, we describe the observations, 
data reduction and the method of the analysis used to characterize 
the host properties. Results and discussion are presented in section 3 and 
the main conclusions drawn in section 4. Throughout this paper, 
to facilitate comparison with literature studies, 
H$_{0}$ = 50 km s$^{-1}$ Mpc$^{-1}$ and q$_{0}$ = 0 are used. 

\section{Observations, data reduction and modeling of the luminosity profiles}

The observations were carried out at the 2.5m Nordic Optical Telescope 
(NOT) in July 2002 (visitor mode) and August 2003 (service mode), 
using the 1024x1024px NOTCam NIR camera with pixel scale 
0.235$''$ px$^{-1}$, giving a field of view of $\sim$4x4 arcmin$^2$. 
The $Ks$-band (2.1 $\mu$m), corresponding to 
$\sim$0.9-1.3 $\mu$m rest-frame, i.e. the minimum in 
the nucleus/host luminosity ratio, was used for all the observations. 
The seeing during the observations was generally very good, 
ranging from 0\farcs55 to 1\farcs05 arcsec FWHM (average and median 
0\farcs7 FWHM). A journal of the  observations is given in 
Table ~\ref{journal}. For all these distant targets, the images were 
acquired keeping the target in the field by dithering it across the array in 
a random grid with typical offsets of $\sim$20 arcsec and acquiring 
several short (60 sec) exposures at each position. Individual exposures 
were then coadded to achieve the final integration time 
(Table ~\ref{journal}).

Data reduction was performed using the NOAO Image Reduction and 
Analysis Facility (IRAF\footnote{IRAF is distributed by 
the National Optical Astronomy Observatories, which are operated by 
the Association of Universities for Research in Astronomy, Inc., 
under cooperative agreement with the National Science Foundation.}). 
Bad pixels were identified via a mask made from the ratio of two 
sky flats with different illumination level, and were substituted 
by interpolating the signal across neighboring pixels. For each 
science image sky subtraction was obtained using a median averaged frame 
of all the other temporally close frames in a grid of eight exposures, 
after scaling it to match the median intensity level of individual 
science frames. Flat-fielding was made using normalized 
median averaged twilight sky frames with different illumination level, 
and images of the same target were aligned and combined using field stars 
as reference points to obtain the final reduced co-added image. 

Observations of standard stars taken from Hunt et al. (1998) were 
obtained throughout the nights in order to provide 
photometric calibration which resulted in accuracy $\sim$0.1 mag from 
the comparison of all observed standard stars. K--correction 
was applied to the host galaxy magnitudes following the method of 
Neugebauer et al. (1985). The size of this correction is significant at 
high redshift in the $K$-band (m(K) = -0.68 and m(K) = -0.59 at z = 0.5 and 
z = 1.0, respectively). No K--correction was applied to 
the nuclear component, assumed to have a power-law spectrum 
($f_\nu \propto \nu^{-\alpha}$) with $\alpha$ $\sim$--1. 
Interstellar extinction corrections were computed using $R$-band 
extinction coefficient from Urry et al. (2000) and A$_K$/A$_R$ = 0.150 
(Cardelli, Clayton \& Mathis 1989).

\begin{table*}
\caption{Previous NIR photometry of high redshift BL Lacs.$^{\mathrm{a}}$}
\label{nirphot}
\begin{flushleft}
\begin{tabular}{llll}
\hline
\noalign{\smallskip}
Name & K mag & K mag range & Refs\\
& this work & literature & \\
(1) & (2) & (3) & (4) \\
\noalign{\smallskip}
\hline
\noalign{\smallskip}
PKS 0138-097   & 12.8 & 12.7 - 14.3 & A82 M90\\
PKS 0235+164   & 12.8 & 8.8 - 14.3 & A82 F99 G85 G86 G93 H84 I82 I84 M90 O78\\
B2 1308+326    & 12.5 & 10.6 - 15.0 & F99 G85 G86 G93 H84 I82 I84 O78 \\
PKS 1538+149   & 13.6 & 13.2 - 14.6 & A82 F99 G93 I82 \\
S5 1803+784    & 12.1 & 12.4 & F99\\
S4 1823+568    & 12.8 & 13.0 - 13.1 & G93\\
PKS 2131-021   & 14.2 & 14.2 - 14.9 & A82 G93\\
PKS 2207+020    & 14.9 & 15.4 & A82\\
\noalign{\smallskip}
\hline
\end{tabular}
\end{flushleft}
\begin{list}{}{}
\item[$^{\mathrm{a}}$] 
Column (1) gives the name of the object; (2) the $K$-band magnitude in 
this work; (3) the range of $K$-band magnitudes in literature; 
(4) references for literature values: A82 = Allen, Ward \& Hyland (1982); 
F99 = Fan \& Lin (1999); G85 = Gear et al. (1985); G86 = Gear et al. (1986); 
G93 = Gear (1993); H84 = Holmes et al. (1984); I82 = Impey et al. (1982); 
I84 = Impey et al. (1984); 
M90 = Mead et al. (1990); O78 = O'Dell et al. (1978).
\end{list}
\end{table*}

For eight sources we found published NIR photometry in the literature 
(Table ~\ref{nirphot}). Our $K$-band photometry agrees generally well with 
the previous studies, considering their intrinsic flux variability. 
In five cases, our new NIR photometry is within the literature range, 
while for three sources, our photometry is slightly brighter than in 
the literature. 

In order to characterize the properties of the host galaxies, 
azimuthally averaged one-dimensional radial luminosity profiles were 
extracted for each BL Lac object and for field stars in the frames 
out to surface brightness $\mu$(K) $\sim$23 mag arcsec$^{-2}$, 
depending on exposure time and observing conditions. All regions affected 
by nearby companions were masked out from the profiles. In most cases, 
there were several field stars available, and the core and the wing of 
the PSF were derived from faint and bright field stars, respectively. 
When available, the profiles of several faint stars were combined to 
represent the PSF wing. Comparison of the final PSF and the profiles 
of individual stars in the frames, as well as comparison between the PSFs 
for each target, indicates that this procedure resulted in a good and 
stable representation of the true PSF. The only exception was the field of 
1ES 1533+535, where only a few faint stars were available. In this case, 
the PSF core was estimated from these stars, whereas the PSF wing was 
derived from the fields of PKS 0138-097 and PKS 2032+107 that were 
observed during the same night and with similar seeing and 
airmass conditions.  

The luminosity profiles were decomposed into a point source (modeled by 
a PSF) and galaxy components (convolved with the PSF) by an 
iterative least-squares fit to the observed profile. There are three 
free parameters in the fit: the PSF normalization, 
the host galaxy normalization, and the effective radius of the host galaxy. 
We attempted both elliptical (de Vaucouleurs r$^{1/4}$ law) and 
exponential r$^{-1}$ disc models to represent the host galaxy. 
The host galaxy was deemed to be resolved, if the best PSF + host galaxy 
fit resulted in significantly lower reduced $\chi^2$ value than 
the PSF-only fit. Consistently with the results from previous work in 
the field but for objects at lower redshift (see references above), 
in no case did the disc model give a significantly better fit than 
the elliptical one. In the following, therefore, we consider only 
the results from the fit with the elliptical model. 

\begin{table*}
\caption{Properties of the host galaxies.$^{\mathrm{a}}$}
\label{hostprop}
\begin{flushleft}
\begin{tabular}{llllllllrllrll}
\hline
\noalign{\smallskip}
& z & A(K)$^{\mathrm{b}}$ & K-corr & m$_K$(n) & m$_K$(g) & $\mu_e$$^{\mathrm{c}}$ & r$_e$ & R$_e$ & M$_K$(n)$^{\mathrm{d}}$ & M$_K$(g)$^{\mathrm{e}}$ & $\pm$ & N/G & Note\\
& & mag & mag & & & & arcsec & kpc & & & & & \\
(1) & (2) & (3) & (4) & (5) & (6) & (7) & (8) & (9) & (10) & (11) & (12) & (13) & (14)\\
\noalign{\smallskip}
\hline
\noalign{\smallskip}
PKS 0138-097   & 0.733 & 0.02 & -0.66 & 13.25 & 16.17 & 17.58 & 0.41 & 3.9  & -30.77 & -27.21 & 0.43 & 26.5 & M\\
PKS 0235+164   & 0.940 & 0.02 & -0.61 & 12.97 & $>$ 13.95 &   &      &   & -31.55 & $>$ -30.16 & & $>$ 3.6 & U\\
B2 1308+326    & 0.997 & 0.01 & -0.59 & 12.95 & 15.49 & 18.18 & 0.72 & 7.8  & -31.82 & -28.69 & 0.39 & 17.9 & M\\
RXJ 14226+5801 & 0.638 & 0.01 & -0.67 & 15.61 & 16.20 & 18.83 & 0.70 & 6.4  & -27.92 & -26.66 & 0.31 & 3.2 & R\\
1ES 1517+656   & 0.702 & 0.02 & -0.67 & 13.79 & 14.96 & 18.12 & 0.90 & 8.6  & -30.01 & -28.17 & 0.31 & 5.4 & R \\
1ES 1533+535   & 0.890 & 0.01 & -0.62 & 16.67 & 16.78 & 17.93 & 0.48 & 5.0  & -28.07 & -27.35 & 0.20 & 1.9 & R \\
PKS 1538+149   & 0.605 & 0.04 & -0.67 & 14.28 & 15.31 & 19.93 & 1.75 & 15.6 & -29.13 & -27.43 & 0.40 & 4.8 & R \\
S4 1749+701    & 0.770 & 0.09 & -0.66 & 13.70 & 15.42 & 16.83 & 0.40 & 4.0  & -30.42 & -28.04 & 0.34 & 9.0 & R \\
S5 1803+784    & 0.684 & 0.04 & -0.67 & 12.61 & 14.14 & 17.03 & 0.79 & 7.4  & -31.14 & -28.93 & 0.20 & 7.7 & R \\
S4 1823+568    & 0.664 & 0.04 & -0.67 & 13.41 & 14.68 & 16.58 & 0.50 & 4.6  & -30.25 & -28.31 & 0.21 & 6.0 & R \\
PKS 2032+107    & 0.601 & 0.04 & -0.67 & 12.35 & 15.50 & 18.28 & 0.75 & 6.6 & -31.05 & -27.23 & 0.59 & 33.7 & M \\
PKS 2131-021   & 1.285 & 0.04 & -0.53 & 14.70 & 16.52 & 17.93 & 0.40 & 4.7  & -30.85 & -28.51 & 0.29 & 8.6 & R \\
PKS 2207+020   & 0.976 & 0.04 & -0.60 & 15.20 & $>$ 17.03 &   &      & & -29.31 & $>$ -27.19 & & $>$ 7.0 & U \\
\noalign{\smallskip}
\hline
\end{tabular}
\end{flushleft}
\begin{list}{}{}
\item[$^{\mathrm{a}}$] 
Column (1) and (2) give the name and redshift of the object; 
(3) the interstellar extinction correction; (4) the K--correction for 
first-ranked elliptical galaxies from Neugebauer et al. (1985), 
interpolated to the redshifts of the BL Lacs; (5) and (6) the apparent nuclear 
and host galaxy magnitude; (7) the surface brightness $\mu$(e); 
(8) and (9) the bulge scalelength in arcsec and kpc; 
(10) and (11) the absolute nuclear and host galaxy magnitude; 
(12) the $\pm$ error estimate of the derived host galaxy luminosity 
(see text for details); (13) the nucleus/host galaxy luminosity ratio; 
and (14) R = resolved; M = marginally resolved; U = unresolved. 
\item[$^{\mathrm{b}}$] 
The interstellar extinction corrections were computed using 
$R$-band extinction coefficient from Urry et al. (2000) and 
A$_K$/A$_R$ = 0.150 (Cardelli et al. 1989).
\item[$^{\mathrm{c}}$] 
Corrected for Galactic extinction, K--correction and cosmological dimming.
\item[$^{\mathrm{d}}$] 
Corrected for extinction only. The BL Lac nuclei are assumed to have flat 
power law spectra and therefore have negligible K--correction.
\item[$^{\mathrm{e}}$] 
Corrected for extinction and K--correction.
\end{list}
\end{table*}

The uncertainty of the derived host galaxy magnitudes depends mainly on 
the nucleus/host luminosity ratio. The uncertainty of the host magnitude for 
each object has been evaluated from the variation of the two relevant 
parameters ($\mu_o$ and r$_e$) from their best-fit value within 
the $\chi^2$ map and assuming a $\Delta\chi^2$ = 2.7 (e.g. Avni 1976). 
These uncertainties are given in Table ~\ref{hostprop}, column 12. 
The estimated uncertainty in the host galaxy luminosity ranges from 
$\sim$$\pm$0.2 mag to $\sim$$\pm$0.6 mag, with the three marginally resolved 
hosts having the largest uncertainty. The average value for the sample is 
$\sim$$\pm$0.3 mag. 

\section{Results}

\begin{figure*}
\centering
\includegraphics[width=15cm]{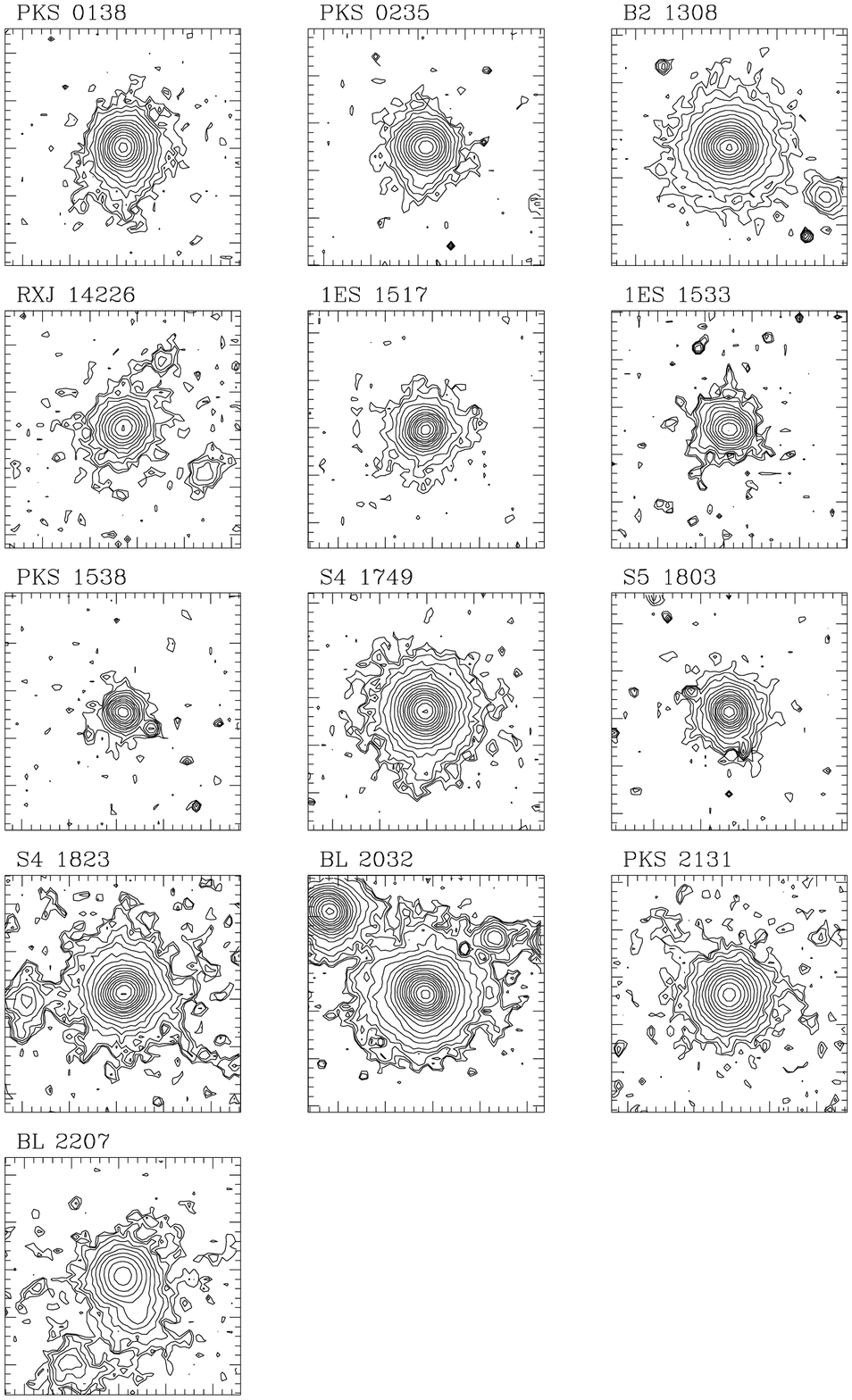}
\caption{Contour plots of the BL Lac objects (center) in the $K$-band. 
The distance between major tick marks is 10 px (2\farcs35). 
Successive isophotes are separated by 0.5 mag intervals. North is up and 
east to the left.
\label{fig1}}
\end{figure*}

\begin{figure*}
\centering
\includegraphics[width=15cm]{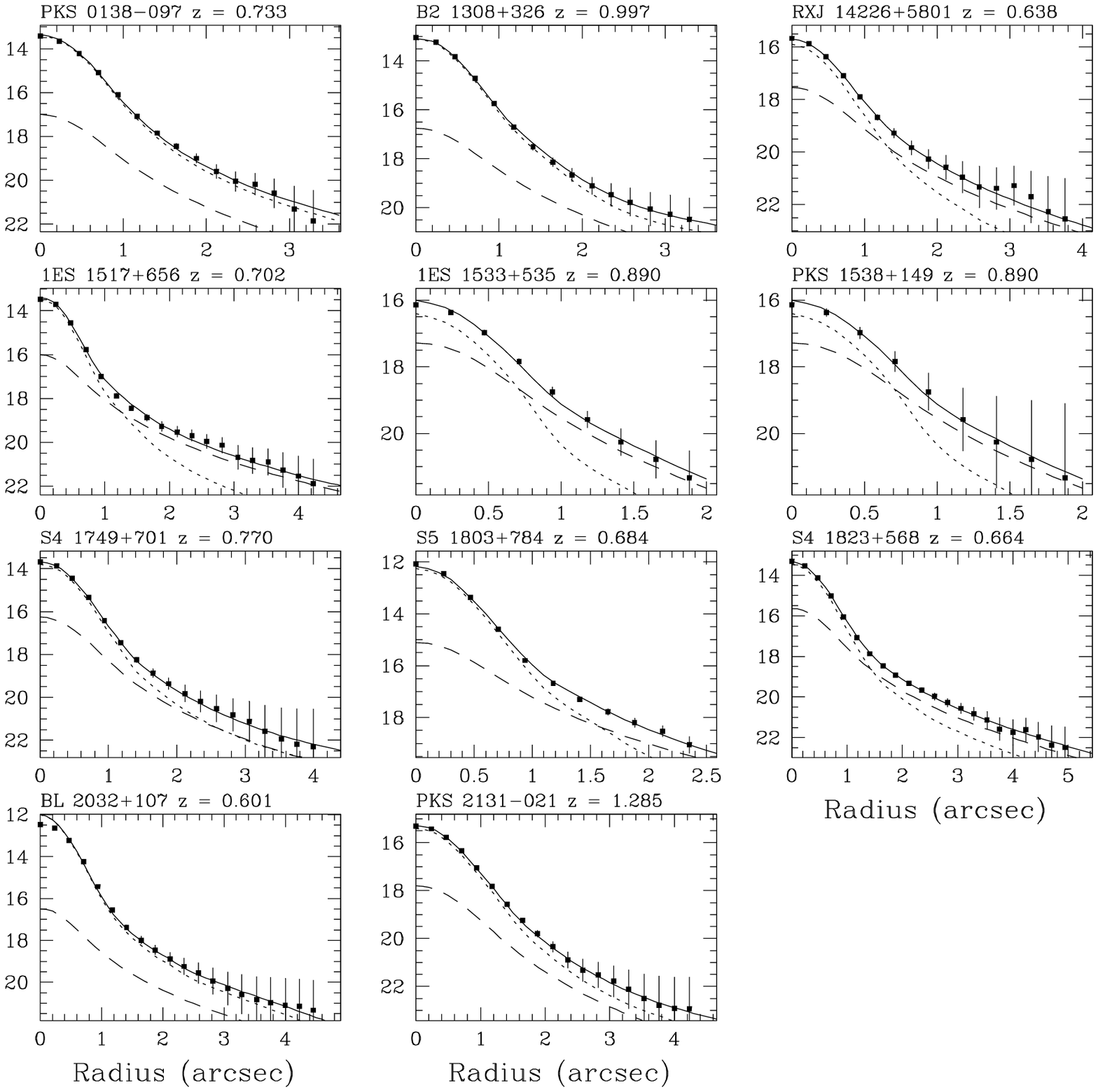}
\caption{The observed $K$-band azimuthally averaged radial surface 
brightness profile (solid points with error bars) for each BL Lac object, 
overlaid with the scaled PSF model (dotted line), the de Vaucouleurs $r^{1/4}$ 
model convolved with the PSF (long-dashed line), and the fitted PSF + 
host galaxy model profile (solid line). The y-axis is in mag arcsec$^{-2}$.
\label{fig2}}
\end{figure*}

In Fig.~\ref{fig1} we show the $K$-band contour plots of all the observed 
BL Lacs, We are able to clearly detect the host galaxy in eight BL Lacs, 
and marginally in three others (PKS 0138-097, B2 1308+326 and 
PKS 2032+107). The host remains unresolved in only two cases contaminated 
by overlapping or nearby sources, PKS 0235+164 and PKS 2207+020 
(see Appendix). This high detection rate is most likely due to 
a combination of filter choice (rest-frame $\sim$1 $\mu$m, 
targeting the dominant stellar population in the galaxies), 
relatively long integration times, and subarcsec seeing conditions. 
In Fig.~\ref{fig2}, we show the $K$-band azimuthally averaged 
radial luminosity profiles of each BL Lac object, together with 
the best--fit PSF + elliptical host models overlaid. In the Appendix, 
we discuss in detail the results for individual BL Lacs, 
including comparison with previous optical characterization of 
the host galaxy properties. The results derived from the best-fit 
model parameters (PSF + elliptical host galaxy) of the profile fitting 
are summarized in Table ~\ref{hostprop}.
To establish upper limits to the undetected hosts, we have assumed that 
the size of the host galaxy in the two unresolved BL Lacs corresponds to 
the average value of the sample (R(e) = 7 kpc). We gradually increased 
the luminosity of the host in the fit (by increasing its surface brightness) 
until it would become detectable by our observation within 
the associated errors. The upper limit to the host galaxy magnitude was then 
derived by integrating the profile of the model host galaxy. The derived 
upper limits to the host luminosities are M(K) = -30.2 
for PKS 0235+164 and M(K) = -27.2 for PKS 2207+020. The faint upper limit for 
PKS 2207+020 indicates that the host must belong to the faintest among 
the sample.

\subsection{Luminosity of the host galaxies}

\begin{figure*}
\centering
\includegraphics[width=15cm]{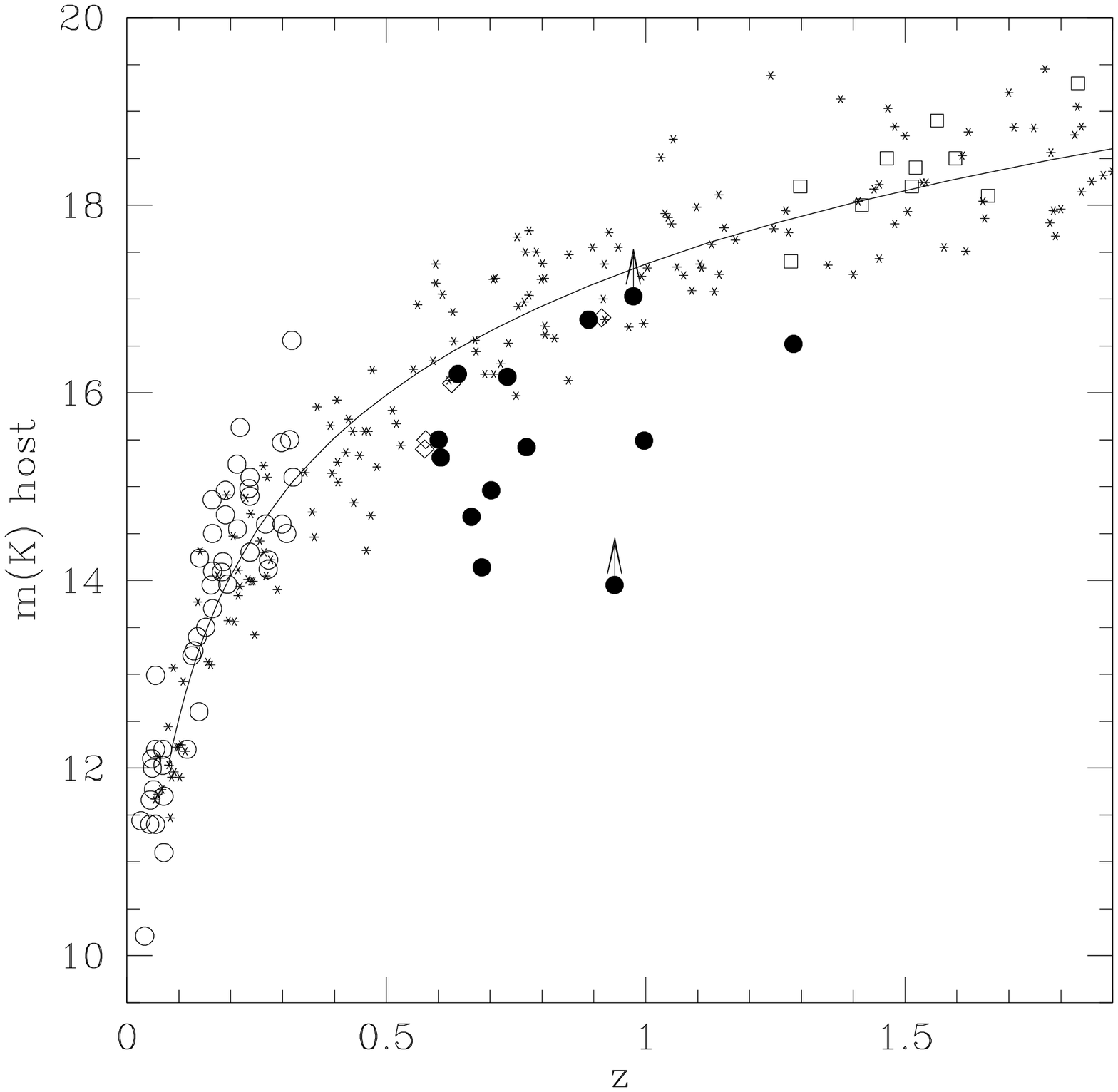}
\caption{
Plot of the apparent $K$-band magnitude of the host galaxies 
vs. redshift (Hubble diagram). The high redshift BL Lacs from this work 
are marked as filled circles, low redshift BL Lcs as open circles 
(Kotilainen et al. 1998a; Scarpa et al. 2000b; Cheung et al. 2003; 
Kotilainen \& Falomo 2004), FSRQ hosts as open diamonds 
(Kotilainen et al. 1998b) and high redshift quasar hosts as open squares 
(Falomo et al. 2004). The solid line is the $K$-z relation for RGs derived 
by Willott et al. (2003). These RGs are shown as asterisks. Filter conversion 
assumes H-K = 0.2.
\label{mhz}}
\end{figure*}

It is now well established that RGs exhibit a smooth apparent magnitude 
vs. redshift ($K$--z) Hubble relation with relatively small scatter out to 
at least z $\sim$3 (e.g. Lilly \& Longair 1984; Eales et al. 1997; 
Jarvis et al. 2001), consistent with a coeval, passively evolving 
galaxy population that formed at z $\geq$4. In Fig.~\ref{mhz}, we show 
the $K$-band Hubble diagram for AGN hosts, namely high redshift 
BL Lac hosts (this work), low redshift BL Lac hosts 
(KFS98, S00, C03 and KF04), FSRQ hosts (Kotilainen et al. 1998a), 
and RLQ hosts (Falomo et al. 2004), compared to the most recent 
relation obtained for RGs (solid line; Willott et al. 2003). Note that 
the majority of the low redshift BL Lacs lie on or above the established 
RG relation, i.e. toward fainter magnitude by $\sim0.5$ mag on the average. 
On the other hand, the high redshift BL Lac hosts and the firmly detected 
FSRQ hosts tend to be displaced in the opposite direction, 
i.e. toward brighter magnitude, suggesting that at around z $\sim$0.5 
there is an increase in the host brightness with respect to 
the Hubble diagram defined by the RGs. Interestingly, this increase is 
not followed by the hosts of more powerful radio-loud AGN 
(Falomo et al. 2004), which are in good agreement with the Hubble diagram. 

\begin{figure*}
\centering
\includegraphics[width=15cm]{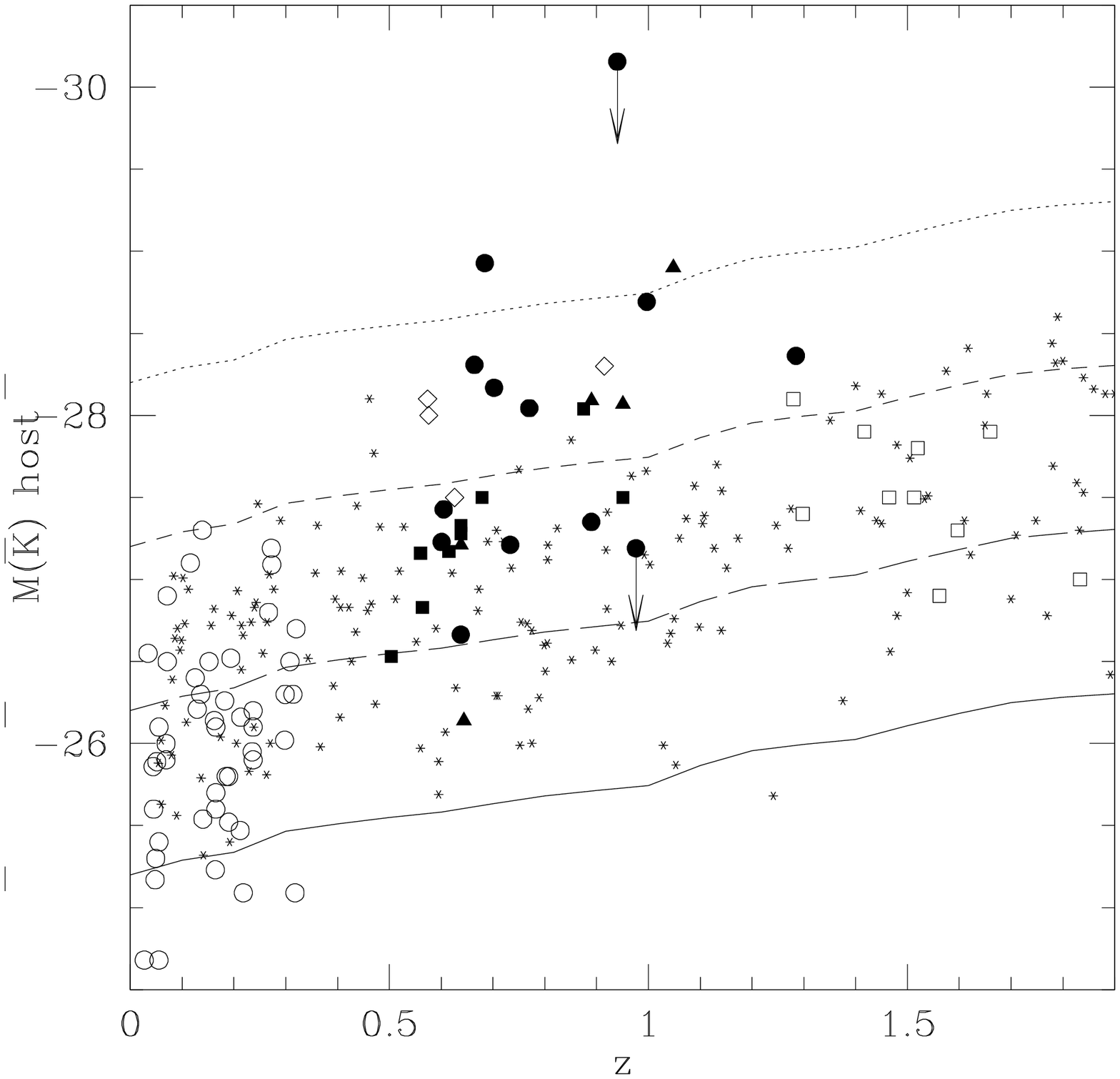}
\caption{
Plot of the absolute $K$-band magnitude of the host galaxies 
vs. redshift. For symbols, see Fig.~\ref{mhz}. Additional symbols indicate 
optical studies of high redshift BL Lacs (filled squares; Heidt et al. 2004, 
and filled triangles; O'Dowd \& Urry 2004). Filter conversions assume 
R-K = 2.7; I-K = 2.0 and H-K = 0.2. The solid, long-dashed, 
short-dashed and dotted lines are the luminosities of L* (M(K)$\sim$-25.2 
at low redshift; Mobasher et al. 1993), L*-1, L*-2 and L*-3 galaxies, 
respectively, following the passive evolution model of 
Bressan, Granato \& Silva (1998).
\label{Mhz}}
\end{figure*}

In Fig.~\ref{Mhz}, we show the $K$-band host galaxy absolute magnitude 
vs. redshift for the BL Lacs (this work, KFS98, S00, C03, KF04, OU04 
and H04), FSRQs (Kotilainen et al. 1998a), RLQs (Falomo et al. 2004) 
and RGs (Willott et al. 2003). The average 
$K$-band absolute magnitude of the high redshift (z$\geq$0.5) BL Lac hosts 
in this study is $<$M(K)$>$ = -27.9$\pm$0.7. 
This is $\sim$2.5 mag 
more luminous than the characteristic magnitude for nearby 
non-active ellipticals (an L$^*$ galaxy with 
$<$M(K)$>$ = --25.2$\pm$0.3; Mobasher, Sharples \& Ellis 1993), 
and is encompassed 
within M*-1 and M*-3. The BL Lac hosts are, therefore, 
preferentially selected from the high--luminosity tail of 
the galaxy luminosity function. 

\begin{table*}
\caption{Comparison of average host galaxy properties with other samples.$^{\mathrm{a}}$}
\label{samples}
\begin{flushleft}
\begin{tabular}{llrlll}
\hline
\noalign{\smallskip}
Sample & filter & N & $<z>$ & $<M_K(nuc)>$ & $<M_K(host)$ \\
(1) & (2) & (3) & (4) & (5) & (6) \\
\noalign{\smallskip}
\hline
\noalign{\smallskip}
BL (this work) & K & 12 & 0.795$\pm$0.197 & -30.1$\pm$1.2 & -27.9$\pm$0.7\\
BL H04  & I   & 9 & 0.669$\pm$0.140 & & -27.2$\pm$0.8\\
BL OU04 & R   & 5 & 0.843$\pm$0.156 & & -27.7$\pm$0.9\\
	     &	 &   &                 &               &             \\
BL KF04 & H   & 23 & 0.155$\pm$0.091 & -24.5$\pm$1.6 & -26.0$\pm$0.7 \\
BL C03  & K   & 8  & 0.186$\pm$0.088 & -25.6$\pm$0.8 & -26.0$\pm$0.4 \\
BL S00  & H   & 9  & 0.206$\pm$0.074 & -25.0$\pm$1.6 & -26.4$\pm$0.4 \\
BL KFS98 & H  & 7  & 0.112$\pm$0.068 & -25.7$\pm$1.7 & -26.0$\pm$0.5 \\
BL all low redshift & H/K & 42 & 0.164$\pm$0.084 & -25.0$\pm$1.5 & -26.1$\pm$0.6\\
	     &	 &   &                 &               &             \\
L* Mobasher et al. (1993) & K & 136 & 0.077$\pm$0.030 & & -25.2$\pm$0.2 \\
	     &	 &   &                 &               &             \\
BCG Thuan \& Puschell (1989) & H & 84 & 0.074$\pm$0.026 & & -26.5$\pm$0.3 \\
BCG Aragon-Salamanca et al. (1998) & K & 25 & 0.449$\pm$0.266 & & -27.2$\pm$0.3\\
	&	&   &       &         &             \\
RG F--R II Taylor et al. (1996) & K & 12 & 0.214$\pm$0.049 & -25.1$\pm$0.7 & -26.3$\pm$0.8 \\
RG Willott et al. (2003) z $<$ 0.3 & K & 42 & 0.170$\pm$0.075 & & -26.2$\pm$0.7 \\
RG Willott et al. (2003) 0.5 $<$ z $<$ 1 & K & 45 & 0.749$\pm$0.132 & & -26.8$\pm$0.5 \\
RG Willott et al. (2003) 1 $<$ z $<$ 2 & K & 63 & 1.480$\pm$0.306 & & -27.4$\pm$0.7 \\
	&	&   &       &         &             \\
FSRQ/R+M Kotilainen et al. (1998b) & H & 9 & 0.671$\pm$0.157 & -29.7$\pm$0.8 & -26.9$\pm$1.2 \\
FSRQ/R Kotilainen et al. (1998b) & H & 4 & 0.673$\pm$0.141 & -30.2$\pm$0.7 & -28.0$\pm$0.3 \\
SSRQ/R+M Kotilainen \& Falomo (2000) & H & 16 & 0.690$\pm$0.088 & -28.3$\pm$1.3 & -27.2$\pm$1.2 \\
SSRQ/R Kotilainen \& Falomo (2000) & H & 10 & 0.683$\pm$0.077 & -28.2$\pm$1.2 & -27.4$\pm$1.1 \\
	&	&   &       &         &             \\
RLQ Falomo et al. (2004) & H/K & 10 & 1.514$\pm$0.157 & -30.9$\pm$0.8 & -27.5$\pm$0.4 \\
\noalign{\smallskip}
\hline
\end{tabular}
\end{flushleft}
\begin{list}{}{}
\item[$^{\mathrm{a}}$] 
Column (1) gives the sample; (2) the filter; (3) the number of objects in 
the sample; (4) the average redshift of the sample; and (5) and (6) the 
average $K$-band nuclear and host galaxy absolute magnitude of the sample. 
\end{list}
\end{table*}

It is interesting to compare the $K$-band absolute magnitude distribution for 
our BL Lac hosts to those of other relevant samples of BL Lacs, FSRQs and RGs 
from previous NIR studies (see references above). These samples span 
a moderately large range in redshift from z$\sim$0.03 up to z$\sim$1. 
The average host galaxy magnitudes for the various samples are given in 
Table ~\ref{samples}. All published magnitudes were transformed into 
our adopted cosmology (note that OU04 use H$_{0}$ = 70 km s$^{-1}$ Mpc$^{-1}$, 
$\Omega_M$ = 0.3 and $\Omega_\lambda$ = 0.7) and into the $K$-band, 
and based on the similarity of the BL Lac host galaxies to 
giant ellipticals, assuming the average rest-frame colour of 
giant ellipticals, $R$-$K$ = 2.7, $I$-$K$ = 2.0 (Kotilainen et al. 1998a) 
and $H$--$K$ = 0.22 (Recillas-Cruz et al. 1990). 

The hosts of high redshift BL Lacs appear also to be more luminous 
in the NIR than nearby BCGs ($<$M(K)$>$ = --26.5$\pm$0.3; 
Thuan \& Puschell 1989) but only slightly more luminous than
BCGs at higher redshift (z $\sim$0.4; 
$<$M(K)$>$ = -27.2$\pm$0.3; Aragon-Salamanca et al. 1998), 
with many of the BL Lac hosts falling into the BCG range. They are 
also brighter than RGs at 0.5 $<$ z $<$ 1 
($<$M(K)$>$ = -26.8$\pm$0.5; Willott et al. 2003), 
but consistent with RGs at z $>$ 1 ($<$M(K)$>$ = -27.4$\pm$0.7; 
Willott et al. 2003). The BL Lacs in our study also have 
host galaxies significantly brighter than those found by previous 
NIR studies of lower redshift (z $\sim$0.15) BL Lacs 
($<$M(K)$>$ = -26.1$\pm$0.6; KF04). 

On the other hand, the high redshift BL Lac hosts are in good agreement 
with the previous optical studies of high redshift BL Lacs 
($<$M(I)$>$ = -25.2$\pm$0.8; corresponding to 
$<$M(K)$>$ = -27.2$\pm$0.8 [H04] and $<$M(R)$>$ = -25.0$\pm$0.9; 
corresponding to $<$M(K)$>$ = -27.7$\pm$0.9 [OU04]). As noted above, 
we have assumed normal elliptical galaxy colours in converting 
the optical magnitudes into the $K$-band, but this appears reasonable, 
as the small amount of available data indicates that the host galaxy colour 
of high redshift BL Lacs is on average similar to that of 
normal ellipticals (see Appendix). BL Lacs share many properties 
(e.g. variability and polarization) with FSRQs and it is therefore 
also interesting to compare the host properties of the two types  
of blazars. The average host magnitude of the high redshift BL Lac hosts 
is closely similar to the value found for resolved FSRQ hosts at 
0.5$<$z$<$1.0 ($<$M(K)$>$ = --28.0$\pm$0.3; Kotilainen et al. 1998a),

\subsection{Evolution and selection effects}

The main result derived from the above comparison is that high redshift 
BL Lac and FSRQ host galaxies are  very luminous, much brighter than 
those seen at low redshift. They appear also more luminous than what 
is predicted by models of passive stellar evolution from a high redshift 
formation epoch (Fig.~\ref{Mhz}) or by models of a non-evolving population. 
The $\sim$2 mag difference between the high redshift BL Lac (and FSRQ) hosts 
and the low redshift hosts could be either due to a significant 
host luminosity evolution with redshift, or assuming that the nuclear and 
the host luminosity are correlated, due to selection effects because of 
the relatively bright completeness level of flux limited surveys of BL Lacs. 

\begin{figure*}
\centering
\includegraphics[width=15cm]{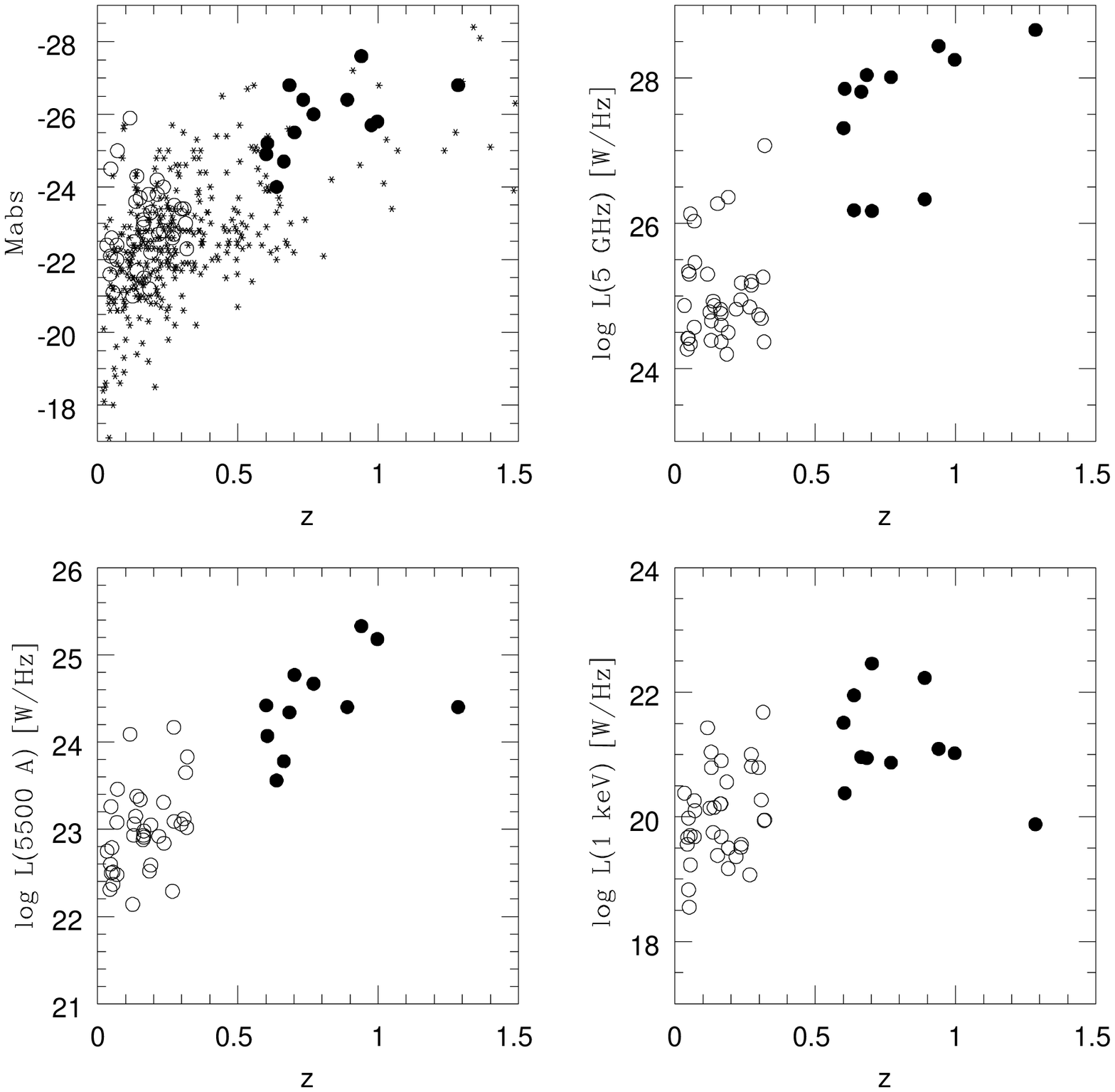}
\caption{
Comparison of high and low redshift BL Lacs (filled and open circles, 
respectively) at various wavelengths. {\bf Top left:} V-band absolute 
magnitude. {\bf Top right:} The 5 GHz radio power. {\bf Bottom left:} 
The 5500 \AA monochromatic luminosity. {\bf Bottom right:} 
The 1 keV luminosity. The first panel uses data from 
Veron-Cetty \& Veron (2003), and the others from Donato et al. (2001).
\label{rox}}
\end{figure*}

In order to investigate the role of the possible selection effects, we have 
analyzed the properties  of the various BL Lac samples in 
the luminosity--redshift plane. Fig.\ref{rox} compares the distribution of 
the high and low redshift BL Lacs in the luminosity-redshift plane at 
radio (5 GHz), optical (V-band and 5500 \AA) and X-ray (1 keV) frequencies. 
It turns out that the nuclei of high redshift BL Lacs (this work) 
are much more luminous at all wavelengths from radio to X-rays than 
the nuclei of low redshift BL Lacs (the best overlap of the distributions is 
in X-rays). The well known flux limit effect with high redshift surveys 
implies a dependence of the luminosity on distance. A similar situation is 
found for quasars (e.g. Falomo et al. 2004). However, in the case of BL Lacs, 
this effect appears to be more significant than in the case of quasars. 
This comparison indicates that it is not possible to directly compare 
the properties of objects belonging to the same population at 
different redshift (which could then be different because of evolution). 
On the contrary, comparison of BL Lac objects at significantly 
different redshifts implies having to study objects belonging to 
different regions of the BL Lac luminosity function. 

Because of the shape of the luminosity funcion of BL Lacs, and of 
the flux limits in the surveys, the large majority of the objects at 
high redshift remain undetected and only the most luminous objects are 
observable. Indeed, at z $\geq$ 0.7 only a few very luminous objects belonging 
to the BL Lac class have been discovered. Therefore, to make an unbiased 
comparison of the host properties at different redshifts, one should choose 
samples that  have matched nuclear luminosity. However, this is in practice 
not possible because in the available BL Lac samples there is very little 
overlap in the nuclear luminosity distributions (Fig.\ref{rox}) for objects 
below and above z $\sim$0.5. 

In spite of the strong selection effect described above, the comparison of 
the host properties for objects at different redshift could be unbiased 
unless there is a significant correlation between the nuclear and 
the host luminosity. 
In the case of low redshift (low power) BL Lacs (see next Section), 
the host galaxy luminosity is unrelated to the nuclear luminosity. 
If this were the case also for high redshift BL Lacs, then the brightening of 
their hosts should be due to strong luminosity evolution of similarly 
massive galaxies. However, for our sample (and the other available 
small samples) of high redshift BL Lacs there seems to be a significant 
correlation between the nuclear and the host luminosity.
Therefore, the alternative view that this correlation, combined with 
the selection effect, causes the most luminous BL Lacs (with the most 
massive BHs) to be preferentially found in the (very rare) most luminous 
(massive) galaxies remains valid. Consequently, the difference in 
nuclear luminosity may fully account for the observed difference in 
host luminosity, and no strong luminosity evolution would be required.

\subsection{The size of the host galaxies}

The average effective radius of the 11 resolved high redshift (z$\geq$0.5) 
BL Lac hosts presented in this study is $<$R$_e$$>$ = 6.8 $\pm$3.2 kpc. 
This is in fair agreement with the typical value of 
intermediate redshift FSRQs (Kotilainen et al. 1998a; 
$<$R(e)$>$ = 11.6$\pm$7.6 kpc). Our value also agrees with that found for 
the high redshift BL Lacs studied by H04 ($<$R(e)$>$ = 10.6$\pm$7.3 kpc), 
but contrast with the significantly larger value reported by OU04 
($<$R(e)$>$ = 32$\pm$17 kpc) for five BL Lac hosts. OU04 suggested that 
their large scale lengths indicate that some of the host galaxies may 
have substantial disk components as well as bulges, but this is 
not corroborated by ground-based studies with a larger field-of-view 
(this work; H04). 

It is worth to note that the host galaxy sizes of high redshift BL Lac hosts 
are similar to those of  lower redshift objects studied in the NIR, 
$<$R(e)$>$ = 7.2$\pm$3.6 kpc (KF04 and references therein).  
This indicates that at least up to z $\sim$1 there is no significant change in 
the global structure of the hosts, in agreement with what is found for RLQs and 
FR II RGs (e.g. McLure \& Dunlop 2000; Kukula et al. 2001; 
Falomo et al. 2004). Our  measurements of the scale length of BL Lacs 
extends to higher redshift the finding that the average size 
 of the BL Lac hosts is a factor $\sim$2-4 smaller than 
that of  low redshift RGs studied in the optical by 
Govoni et al. (2000; $<$R(e)$>$ = 16$\pm$10 kpc) and in the NIR by 
Taylor et al. (1996; $<$R(e)$>$ = 26$\pm$16 kpc), and for 
z $\sim$0.5 RGs ($<$R(e)$>$ = 19$\pm$8 kpc; McLure et al. 2004). 

\subsection{Surface brightness - effective radius (Kormendy) relation}

Normal early-type galaxies and spiral bulges define a continuous tight 
relation between the effective radius, $r_e$, and the surface brightness at 
that radius, $\mu_e$ (the Kormendy relation; Kormendy 1977; 
Kormendy \& Djorgovski 1989). This is indeed a 2D projection of 
the 3D Fundamental Plane linking the above parameters to 
the velocity dispersion $\sigma$. This relation is well established for 
early-type galaxies and spiral bulges in nearby clusters 
(e.g. J\"orgensen, Franx \& Kjaergaard 1996) and is closely related to 
the morphological and dynamical structure of galaxies, and to their 
formation processes. 

\begin{figure*}
\centering
\includegraphics[width=15cm]{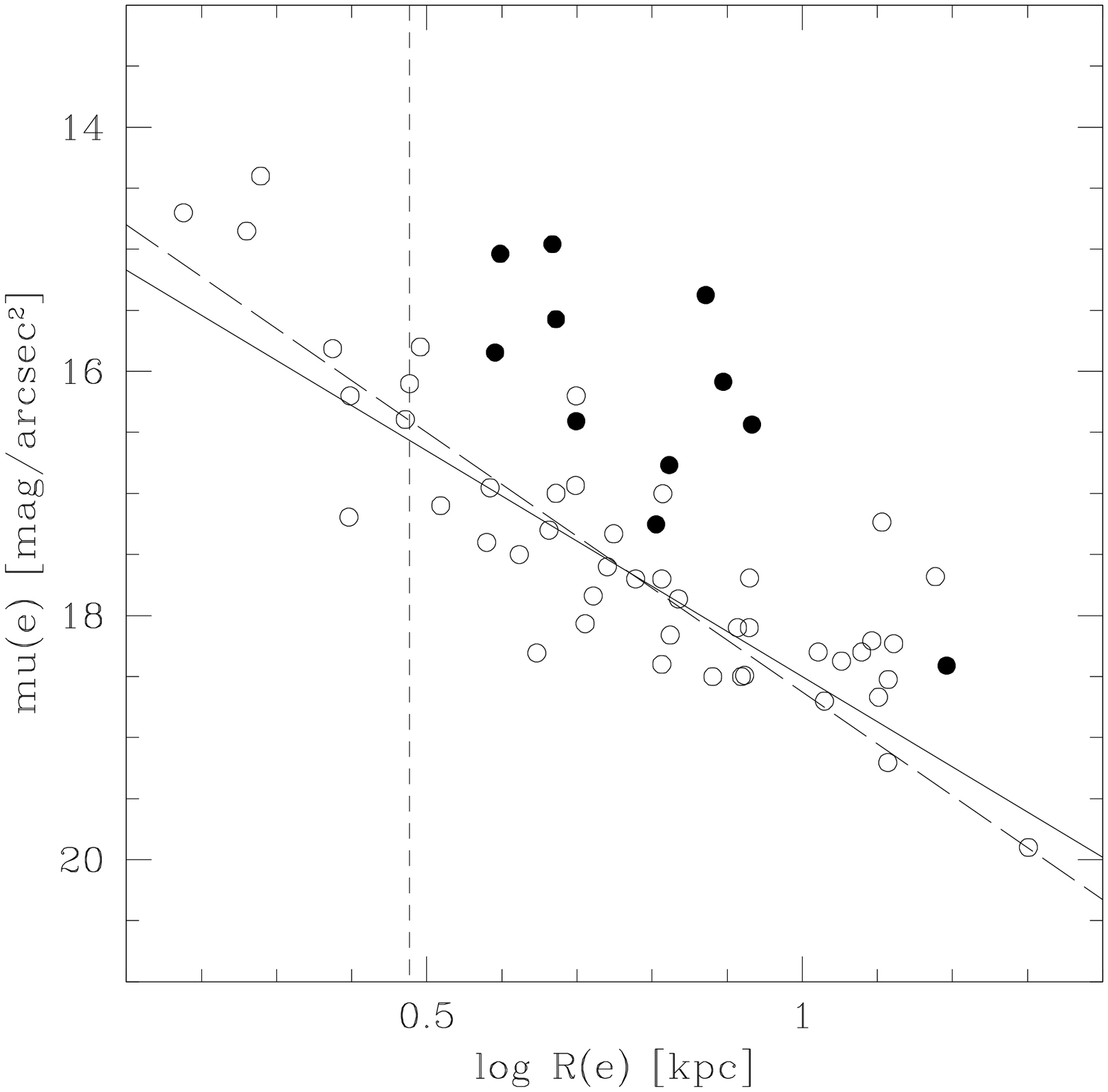}
\caption{
The $K$-band $\mu_e$ - r$_e$ Kormendy relation for the BL Lac host galaxies. 
Conversion from $H$- to $K$- band was made assuming $H-K=0.22$ colour, 
typical of normal ellipticals (Recillas-Cruz et al. 1990). 
Passive evolution of a single age stellar population with formation 
redshift z = 2 in the K-band was adapted from van Dokkum \& Franx (2001).
For symbols, see Fig.~\ref{mhz}. 
The solid and long-dashed lines are linear least-square 
best-fit relations for low redshift BL Lacs (KF04) and normal 
inactive ellipticals (Pahre et al. 1995), respectively. The short-dashed 
vertical line is the dividing line between normal and giant ellipticals at 
R(e) = 3 kpc (Capaccioli et al. 1992). 
\label{muere}}
\end{figure*}

We combined our $\mu_{e}-r_{e}$ data with previous data from KFS98, S00, 
C03 and KF04 to construct the Kormendy relation for the largest available 
sample of BL Lac host galaxies studied in the NIR (Fig.~\ref{muere}). 
The surface brightnesses of the hosts have been corrected for 
galactic extinction and cosmological dimming (10 $\times$ log(1+z)). 
The $H$-band data from KFS98, S00 and KF04 were converted to 
the $K$-band assuming $H-K=0.22$ (Recillas-Cruz et al. 1990). 
We have also included a correction for passive evolution of a single age 
stellar population with formation redshift z = 2 in the $K$-band, 
adapted from van Dokkum \& Franx (2001). The size of this correction 
is $\sim$0.45 mag at z = 0.5, increasing to $\sim$0.9 mag at z = 1. 
Even after these corrections, only a few of the high redshift BL Lac hosts 
are readily in agreement with respect to the relations for low redshift 
BL Lacs (KF04) and for normal elliptical galaxies 
(Pahre, Djorgovski \& de Carvalho 1995). 
Most of the high redshift BL Lac sources are offset from these relations 
towards brighter surface brightness by $\sim$1.5 mag on average, This is 
in agreement with the detected difference of their integrated luminosity.

The unavoidable conclusion is that at z $>$ 0.5 the hosts of high redshift 
BL Lacs are structurally similar (they have the same size) as lower redshift 
BL Lac hosts and normal early-tpye galaxies but they have much higher 
surface brightness. This can be interpreted in terms of 
the passive evolution of the stellar populations, 
if they are dominated by an old stellar population 
with a small fraction of the mass involved in a more recent secondary 
star formation episode. Note that similar offsets have been found for 
high redshift (0.8 $<$ z $<$ 1.9) 3CR RGs (Zirm et al. 2003) and for 
z $\sim$0.5 field early-type galaxies (Treu et al. 2001), in the sense that, 
for a given effective radius, the high redshift galaxies are brighter than 
expected from the local relation.

\subsection{Nuclear and host properties}

The absolute magnitude of the nuclear component of the high redshift BL Lacs 
ranges from M(K) $\sim$-28 to M(K) $\sim$-32, with average 
$<$M(K)$>$ = --30.1$\pm$1.2. This is $\sim$5 mag brighter than the average 
value of the nuclear luminosity of low redshift BL Lacs (KF04 and 
references therein). This $\sim$ factor 100 of difference is the 
consequence of the selection effect 
that prevents the detection of faint BL Lac objects with increasing redshift. 
On the other hand it is worth to note that the nuclear luminosity of 
the high redshift BL Lacs 
is very similar to that of high redshift resolved FSRQs (ranging from 
M(K) $\sim$-29 to $\sim$-32, with average $<$M(K)$>$ = --29.7$\pm$0.8; 
Kotilainen et al. 1998a), suggesting that these objects are prefentially 
chosen from the most luminous tail of the luminosity function.

The difference in the strength of 
the nuclear component at low and high redshift is also evident considering 
the nucleus/galaxy luminosity ratio L(nuc)/L(gal) in Table ~\ref{hostprop}. 
None of the low redshift BL Lacs have L(nuc)/L(gal)$\geq$3 (KF04), 
whereas all FSRQs (Kotilainen et al. 1998a) and all but one of 
the high redshift BL Lacs are above this limit. The high nuclear luminosity 
observed in the high redshift sources may be either due to their 
systematically higher intrinsic nuclear luminosity, or to the fact that we are 
preferentially selecting (because of the luminosity bias effect) 
those objects which are more strongly beamed. Although both effects may play 
a role, the former is likely the dominant one since the beaming distribution 
is independent of the redshift. This is consistent with the fact that 
the high redshift host galaxies are more luminous than those at low redshift. 

\begin{figure*}
\centering
\includegraphics[width=15cm]{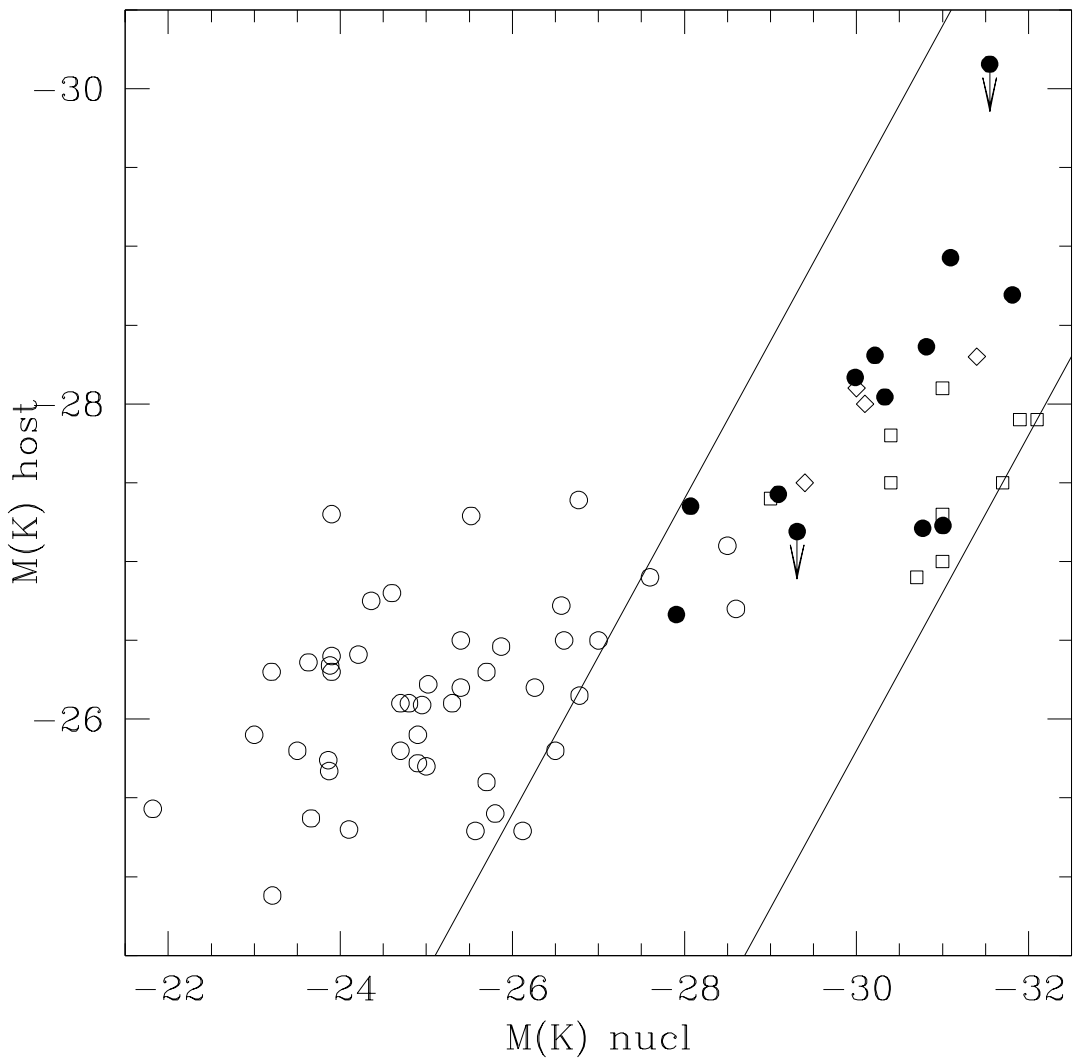}
\caption{
Plot of the $K$-band nuclear vs. host luminosity. For symbols, 
see Fig.~\ref{mhz}. The solid lines represent loci of constant ratio between 
host and nuclear emission. These can be translated into Eddington ratios 
assuming that the central black hole mass - galaxy luminosity correlation 
holds at high redshifts and that the observed nuclear power is proportional to 
the bolometric emission. The two solid lines encompass a spread of 1.5dex in 
the nucleus/host luminosity ratio. 
\label{MhMn}}
\end{figure*}

In Fig.~\ref{MhMn}, we show the relation between the $K$-band luminosities 
of the nucleus and the host galaxy for the BL Lacs, FSRQs and 
high redshift quasars. While the host luminosity of high redshift BL Lacs 
is distributed over a range of only $\sim$2.5 mag ($\sim$4 mag adding 
low redshift BL Lacs), the nuclear luminosities span over $\sim$4 mag 
($\sim$10 mag adding low redshift BL Lacs). While the majority of 
the low redshift (z $<$ 0.5) BL Lac objects are homogeneously distributed 
in the M(nuc)--M(host) plane, the high redshift BL Lacs clearly tend to 
occupy the region of the more luminous host galaxies and nuclei. 
Moreover, they exhibit a reasonable correlation between the host and 
nuclear luminosities, similarly to the resolved FSRQs 
(Kotilainen et al. 1998a). No such correlation is found  for low redshift 
(lower nuclear luminosity) BL Lacs. 

Before discussing the implications of such a relationship, it is worth to 
note that some selection effects may induce a spurious correlation between 
the nuclear and host galaxy luminosity. In particular, two effects may 
combine to depopulate the M(nuc)--M(host) plane in opposite regions. 
Firstly, low luminosity host galaxies are very difficult to detect 
under luminous nuclei. In our sample, however, this should not play 
a significant role because the selection criteria of BL Lacs are only based 
on the nuclear properties, and the objects are not biased with respect 
to their host properties. Indeed, only two host galaxies in our 
sample remained unresolved. Secondly, low luminosity AGN are difficult 
to detect against luminous hosts. This incompleteness may certainly 
be present at high redshift, but this effect is likely to be small 
because high luminosity galaxies are very rare.

The observed relationship between the host and nuclear luminosity, 
therefore, appears genuine and supports the longstanding idea 
(e.g. McLeod \& Rieke 1994) that the onset of the correlation occurs 
only after a certain threshold in the host galaxy luminosity. 
The widely accepted interpretation is that for a given host luminosity, 
the central BH mass is given by the BH mass - bulge mass correlation, 
and that at the highest luminosities (highest BH masses) AGN may accrete 
near their Eddington limit, triggering the onset of 
the BH - host correlation (see also e.g. Lacy, Bunker \& Ridgway 2000; 
Sanchez et al. 2004). Such a correlation appears to follow the trend of 
a roughly constant ratio between the host galaxy luminosity (mass) and 
the nuclear emission. The relatively large ($\sim$1.5dex) scatter would 
then mainly reflect differences in the accretion rates. 

Assuming that the bolometric luminosity scales as the $K$-band luminosity, 
and that the host galaxy luminosity is proportional to the BH mass 
(as seems to be the case for nearby inactive galaxies and 
low redshift AGN; e.g. Marconi \& Hunt 2003; McLure \& Dunlop 2002), 
the nucleus/galaxy luminosity ratio should be proportional to 
the Eddington factor $\xi = L/L_E$, 
where $L_E = 1.25 \times 10^{38} \times (M_{BH}/M_{\odot}$). 
The average $K$-band nucleus-to-host luminosity ratio of the high redshift 
BL Lacs is $<\log(M_{nuc}/M_{host}>$ = 0.90$\pm$0.36. This is about 
a factor of 10 larger than that found for the objects at low redshift  
($<\log(M_{nuc}/M_{host}>$ $\sim$0; Falomo \& Kotilainen 1999; 
Urry et al, 2000; KF04) but still smaller than the ratio found for high redshift RLQs 
($<\log(M_{nuc}/M_{host}>$ = 1.36$\pm$0.33; Falomo et al. 2004). 
These differences are consistent with the idea that BL Lacs are 
intrinsically low power AGN, with much lower accretion rates than those 
in RLQs (as found at low redshift by e.g. O'Dowd et al. 2002). 
The luminous high redshift BL Lacs appear thus, in this context, 
intermediate power AGN between the lower power BL Lacs (which dominate 
the population at low redshift) and the higher power RLQs. 

\section{Summary and conclusions}

We have presented the first near-infrared imaging study of a sizeable sample 
of BL Lac objects at z $>$ 0.5. In 11/13 cases, we are able to resolve 
the objects and to characterize the properties of their host galaxies. 
We find that at z $\sim$0.8 BL Lac host galaxies are very luminous 
(average $<$M(K)$>$ = -27.9$\pm$0.7), $\sim$3 mag brighter than 
L$^*$ galaxies. The BL Lac hosts have similar luminosity to the hosts of 
high redshift FSRQs, consistent with the idea that they form a common class 
of AGN (blazars). The high redshift BL Lac hosts are large 
(average bulge scale length $<$R(e)$>$ = 6.8$\pm$3.2 kpc) and of similar size 
to their low redshift counterparts, indicating that there is no evolution in 
the host galaxy size. 

At variance with the behaviour for low redshift BL Lacs, we find 
a correlation between the luminosity of the host galaxy and that of 
the nucleus in the high redshift BL Lacs. This can occur at 
a fixed fraction of the Eddington luminosity, as the result of a more 
massive BH is interpreted as an increase of emitted power, as expected 
from the BH mass - bulge luminosity correlation found in 
nearby spheroidals. High redshift BL Lacs appear to radiate with 
a wide range of power with respect to their Eddington luminosity, 
and this power is intermediate between the low level observed in nearby 
BL Lacs and the higher level occurring in luminous radio-loud quasars.

The main finding of this work is that the host luminosity is 
$\sim$2 mag brighter than that of lower redshift BL Lacs. This substantial 
increase in luminosity can be due to a combination of a strong 
selection effect in the surveys of BL Lacs that makes observable only 
the most luminous sources at z $>$ 0.5 and produces a correlation between 
the nuclear and the host luminosity that emerges at high redshift.
Alternatively, since this difference of host luminosity is inconsistent with 
a simple passive evolution of the host galaxies, to interpret this difference 
one should invoke a non-negligible contribution from recent star formation 
episodes that takes place at z $>$ 0.5. 

Detailed spectroscopic studies and colour information of these host galaxies 
would clearly help to elucidate this issue. On the other hand, future deeper 
surveys of BL Lacs could provide data to probe the full population of 
BL Lacs up to z $\sim$ 1 and thus significantly reducing the selection effects.

\begin{acknowledgements}
We thank the anonymous referee for helpful suggestions and constructive 
criticism which improved the presentation and interpretation of the results in 
this paper. 
Nordic Optical Telescope is operated on the island of La Palma jointly 
by Denmark, Finland, Iceland, Norway and Sweden, in the Spanish 
Observatorio del Roque de los Muchachos of 
the Instituto de Astrofisica de Canarias. This research has made use of 
the NASA/IPAC Extragalactic Database (NED), which is operated by 
the Jet Propulsion Laboratory, California Institute of Technology, 
under contract with the National Aeronautics and Space Administration. 
JKK and TH acknowledge financial support from the Academy of Finland, 
project 8201017. This work has been partially supported by 
INAF contract 1301/01.
\end{acknowledgements}

\appendix

\section{Appendix: Notes on the host galaxies of individual objects}

{\bf PKS 0138-097}:
This BL Lac object at z = 0.733 has a smooth IR-optical spectrum and is 
highly polarized. Until now, the host galaxy has remained unresolved in 
the optical (e.g. Stickel, Fried \& K\"uhr 1993; Heidt et al. 1996; 
Scarpa et al. 1999, 2000a; Pursimo et al. 2002; OU04) and in the NIR (C03). 
Although Scarpa et al. (1999,2000a) noted a small systematic departure 
from the PSF profile at r $>$ 1 arcsec, 
they concluded that the excess light was due to nearby galaxies,
with upper limit to the host magnitude of m(R) $>$ 20.1 (M(R) $>$ -25.4).
This source was marginally resolved by OU04 but was considered 
unresolved (m(R,host) $>$ 19.6) because the fit was not better than 
the fit with PSF only. The environment near PKS 0138-097 is rich, 
with at least four galaxies within 3 arcsec radius (projected distance of 
30 kpc; Heidt et al. 1996). They could be responsible for 
the intervening absorption system at z = 0.501 
(Stickel et al. 1993; Falomo \& Ulrich 2000). 
The closest of these galaxies (source C in Heidt et al. 1996) is at 
a distance of 1.45 arcsec (projected distance of 14 kpc), 
indicating gravitational interaction with the BL Lac object. The presence 
of absorption systems and close companions led Heidt et al. (1996) 
to suggest that this BL Lac object may be affected by 
gravitational microlensing.
In the NIR, we fit only the profile of the northern part of the source 
(free of intervening galaxies), and the source was found to be 
marginally resolved, with R(e) = 3.9 kpc and M(K,host) = -27.2. 

{\bf PKS 0235+164}:
This BL Lac object at z = 0.940 is well known to be in a complex 
environment and with close companions which in part are responsible for 
the intervening systems at z = 0.852 and z = 0.524 (e.g. Falomo 1996; 
Nilsson et al. 1996 and references therein). Claims of detection of 
a surrounding nebulosity were proposed by Stickel, Fried \& K\"uhr (1988) 
and by 
Abraham et al. (1993) but were not confirmed by later optical studies 
(Falomo 1996; Nilsson et al 1996). The object was also not resolved by HST 
(Scarpa et al. 2000a). Because of the presence of the close companions, 
the extraction of a reliable radial profile is very difficult. 
Instead of masking the regions contaminated by companions, we preferred 
in this case to fit only the part of the image around the target which is 
free of intervening galaxies, and the source was found to be unresolved in 
the NIR. Therefore, PKS 0235+164 has been left out of all discussion of 
the host galaxies. 

{\bf B2 1308+326}:
This object at z = 0.997 has variable emission line strengths 
(Stickel et al. 1993; Scarpa \& Falomo 1997), high bolometric luminosity and 
high Doppler boost factor (Watson et al. 2000), and it may be classified as 
a composite quasar/BL Lac object. It has a number of faint companion galaxies, 
the closest at $\sim$5 arcsec (55 kpc) to SW.
Nilsson et al. (2003) obtained a deep $R$-band exposure (6000 s) under good 
seeing conditions (0.9 arcsec), yet no sign of the host galaxy was seen 
(m(R,host) $>$ 20.6). 
Urry et al. (1999) found the object also unresolved in their deep HST exposure 
(m(I,host) $>$ 20.1). This source was marginally resolved by OU04 but was 
considered unresolved (m(R,host) $>$ 20.1) because the fit was not better than 
the fit with PSF only.
Due to its high redshift, this object has been suggested as a case of 
microlensing by stars in a foreground galaxy.

{\bf RXJ 14226+5801}:
The host galaxy of this BL Lac object at z = 0.638 was not resolved in 
the HST survey (Scarpa et al. 2000a; m(R,host) $>$ 21.7). However, it was 
resolved in the groundbased $I$-band imaging by H04 (r(e) = 2.3 arcsec; 
R(e) = 21 kpc; m(I) = 18.9; M(I) = -25.3). 
The host was also well resolved by OU04, 
with R(e) = 24 kpc and M(R,host) = -24.5. 
We detect the host galaxy in the $K$-band with r(e) = 0.7 arcsec 
(R(e) = 6.4 kpc), i.e. smaller extent than in the optical, and with 
m(K,host) = 16.2 (M(K,host) = -26.7). The integrated colours of 
the host galaxy are thus R-K = 2.2 and I-K = 1.4, slightly bluer than for 
normal ellipticals (R-K = 2.7; I-K = 2.0). 
There are two companion galaxies at $\sim$4 arcsec (36 kpc) to the NW and SW.

{\bf 1ES 1517+656}:
This BL Lac at z = 0.702, although being a high frequency peaked BL Lac, 
is one of the most luminous BL Lac objects in the X-ray, optical and 
radio range. It remained unresolved in the groundbased $R$-band imaging by 
Nilsson et al. (2003), and by Falomo \& Kotilainen (1999).
It was also unresolved in the HST snapshot survey 
(Scarpa et al. 1999, 2000a), with an upper limit m(host) $>$ 19.9, 
corresponding to M(host) $>$ -25.2. 
This BL Lac shows an unusual morphology with three non-homogeneous arclets 
describing an almost perfect Einstein ring surrounding the nucleus at 
2.4 arcsec (23 kpc) distance (Scarpa et al. 1999), 
suggestive of gravitational lensing of one or more background galaxies by a 
foreground galaxy. Deeper images (Falomo \& Kotilainen 1999) show that 
there are many other faint sources in the close environment. They found 
no signature of a massive foreground galaxy, therefore weakening 
the lens hypothesis. In this work, the host galaxy is detected in the NIR, 
with R(e) = 8.6 kpc and M(K,host) = -28.2. 

{\bf 1ES 1533+535}:
The radial profile of this BL Lac at z = 0.890 showed some deviations from 
the PSF model at large radii in the HST survey (Scarpa et al. 2000a), 
but the source was considered unresolved (m(R,host) $>$ 19.7). The host galaxy 
was well resolved by OU04, 
with r(e) = 3\farcs4 (R(e) = 36 kpc) and M(R,host) = -25.4. In the $K$-band, 
we detect the host galaxy with m(K) = 16.5 (M(K) = -27.3) and 
r(e) = 0.4 arcsec (R(e) = 4.2 kpc). The colour of the host galaxy is thus 
R-K = 1.9, bluer than in normal ellipticals. 

{\bf PKS 1538+149}:
PKS 1538+149 is at z = 0.605. 
The host was only barely resolved in the HST snapshot survey by 
Scarpa et al. (2000a; m(R) = 20.2; M(R) = -24.6; r(e) = 2\farcs5; 
R(e) = 22 kpc). However, in deeper HST observations, this object was 
well resolved (Urry et al. 1999; m(I) = 19.0; M(R) = -25.1; r(e) = 2\farcs4; 
R(e) = 21 kpc). 
Nilsson et al. (2003) detected the host with m(R) = 19.9 (M(R) = -24.8) and 
r(e) = 3\farcs0 (R(e) = 27 kpc).
The host galaxy remained unresolved in our previous NIR imaging 
(Kotilainen et al. 1998a), obtained in poor seeing (FWHM $\sim$1\farcs7). 
In this work, we resolve the host in the NIR and derive for it R(e) = 15.6 kpc 
and M(K,host) = -27.4. The NIR host scalelength is the largest for the 
high redshift BL Lac sample, although smaller than the optical scalelength. 
The host galaxy colour is R-K $\sim$2.7, in agreement with normal ellipticals. 
There are several faint galaxies within 50 kpc of the BL Lac 
(Urry et al. 1999), consistent with L* galaxies at the redshift of 
the BL Lac object. 

{\bf S4 1749+701}:
This BL Lac object at z = 0.770 remained unresolved with HST 
(Scarpa et al. 2000a; M(R) $>$ -27.1) and in deeper ground-based 
$R$-band imaging (Nilsson et al. 2003; m(R,host) $>$ 18.7; M(R) $>$ -26.1). 
We detected the host galaxy in the $K$-band with m(K) = 15.4 (M(K) = -28.0).

{\bf S5 1803+784}:
This object (z = 0.684) remained unresolved in the studies by 
Stickel et al. (1993), Scarpa et al. (2000a) and Pursimo et al. (2002). 
For example, Pursimo et al. (2002) derived an upper limit to 
the host luminosity of M(R) $>$ -27.2. We detected the host galaxy in 
the $K$-band with R(e) = 7.4 kpc and M(K) = -28.9).

{\bf S4 1823+568}:
This BL Lac at z = 0.664 was resolved by Nilsson et al. (2003), 
with m(R) = 20.0 (M(R) = -25.2) and r(e) = 2\farcs5 (R(e) = 23 kpc). 
It was also resolved with the HST by Scarpa et al. 2000a (m(R) = 20.2; 
M(R) = -25.1; r(e) = 0\farcs6; R(e) = 5.6 kpc) and by Urry et al. 1999 
(m(I) = 18.8; M(I) = -25.6; r(e) = 1\farcs1; R(e) = 9.3 kpc). 
We detected the host galaxy in the $K$-band with R(e) = 4.6 kpc and 
M(K) = -28.3). This corresponds to a host colour of R-K $\sim$3.2, 
redder than in normal ellipticals. There are several faint galaxies in 
the field, especially 
a nonstellar object 5 arcsec (46 kpc) east of the source. This object is 
highly distorted with asymmetric faint emission elongated toward the south, 
and it may have suffered strong tidal interaction with the host galaxy.

{\bf PKS 2032+107}:
To our knowledge, no previous imaging data exist for this BL Lac object at 
z = 0.601. In this work, the host galaxy is detected in the NIR, with 
R(e) = 6.6 kpc and M(K,host) = -27.2. 

{\bf PKS 2131-021}:
This is the highest redshift BL Lac object in the sample (z = 1.285). It has 
a faint companion galaxy $\sim$4 arcsec (47 kpc) to NE. The host galaxy 
remained unresolved in the optical studies by 
Scarpa et al. (2000a; M(R) $>$ -27.0), H04 (M(I) $>$ -28.3), 
and Pursimo et al. (2002) (M(R) $>$ -27.8). 
Despite the mediocre seeing (FWHM $\sim$1\farcs0), we detected the host galaxy 
in the $K$-band with R(e) = 4.7 kpc and M(K) = -28.5). 

{\bf PKS 2207+020}:
To our knowledge, no previous imaging data exist for this BL Lac object at 
z = 0.976. In the close environment, there are two companion galaxies, 
at 5\farcs1 (55 kpc) SE and at 2\farcs0 (22 kpc) S--SW. The isophotes of 
the latter companion superimposed on the target result in 
a cometary appearance. The radial profile of the target was derived avoiding 
these nearby sources, and was found to be unresolved. 
Therefore, PKS 2207+020 has been left out of all discussion of 
the host galaxies. 

\end{document}